\documentclass[%
 jmp,
 amsmath,amssymb,
 reprint,%
]{revtex4-1}

\usepackage{graphicx}% Include figure files
\usepackage{dcolumn}% Align table columns on decimal point
\usepackage{bm}% bold math
\usepackage[utf8]{inputenc}
\usepackage[T1]{fontenc}
\usepackage{mathptmx}
\usepackage{etoolbox}

\makeatletter
\def\@email#1#2{%
 \endgroup
 \patchcmd{\titleblock@produce}
  {\frontmatter@RRAPformat}
  {\frontmatter@RRAPformat{\produce@RRAP{*#1\href{mailto:#2}{#2}}}\frontmatter@RRAPformat}
  {}{}
}%
\makeatother

\usepackage[normalem]{ulem}
\usepackage{xcolor}
\definecolor{amber}{rgb}{1,0.49,0}
\definecolor{darkgreen}{rgb}{0,0.55,0}
\definecolor{tangerine}{rgb}{0.944,0.522,0}
\definecolor{verde}{rgb}{0.,0.6,0}
\definecolor{rosso}{rgb}{0.9,0.0,0.2}
\definecolor{magenta}{rgb}{0.9,0.2,0.9}

\newif\ifhighlight

\newcommand{\highlight}{\highlighttrue}
\highlight
\newcommand{\editor}[2]{%
  \expandafter\newcommand\csname #1note\endcsname[1]{%
    \textcolor{#2}{(\textbf{#1:} ##1)}}%
  \expandafter\newcommand\csname #1\endcsname[1]{%
    \ifhighlight\textcolor{#2}{##1} \else ##1\fi}%
  \expandafter\newcommand\csname #1cancel\endcsname[1]{%
    \ifhighlight\textcolor{#2}{\sout{##1}}\fi}%
  \expandafter\newcommand\csname #1change\endcsname[2]{%
    \ifhighlight\textcolor{#2}{\sout{##1} ##2}\else ##2\fi}%
  \newenvironment{#1text}{\ifhighlight\color{#2}\fi}{\color{black}}
}

\editor{SB}{tangerine}
\editor{CM}{blue}
\editor{ED}{purple}

\begin{document}
\preprint{AIP/123-QED}

\title[Dynamical Heterogeneity in Supercooled Water and its Spectroscopic
Fingerprints]{Dynamical Heterogeneity in Supercooled Water and its Spectroscopic Fingerprints}
% Force line breaks with \\
\author{Cesare Malosso}
\email{cesare.malosso@epfl.ch}
 \affiliation{Laboratory of Computational Science and Modeling, IMX, École Polytechnique Fédérale de Lausanne, 1015 Lausanne, Switzerland}
 \affiliation{SISSA -- Scuola Internazionale Superiore di Studi Avanzati, Trieste 34136, Italy}%
%Lines break automatically or can be forced with \\
\author{Edward Danquah Donkor}
\affiliation{Department of Applied Science and Technology, Politecnico di Torino, Torino
10129, Italy}
\affiliation{Condensed Matter and Statistical Physics (CMSP), The Abdus Salam Centre for Theoretical Physics, Trieste 34151, Italy}
\author{Stefano Baroni}%
 
\affiliation{SISSA -- Scuola Internazionale Superiore di Studi Avanzati, Trieste 34136, Italy}%
\affiliation{Consiglio Nazionale delle Ricerche -- Istituto Officina dei Materiali, SISSA Unit, Trieste 34136, Italy}

\author{Ali Hassanali}
\affiliation{Condensed Matter and Statistical Physics (CMSP), The Abdus Salam Centre for Theoretical Physics, Trieste 34151, Italy}%

\date{\today}% It is always \today, today,
             %  but any date may be explicitly specified

\begin{abstract}
A growing body of theoretical and experimental evidence strongly supports the existence of a second liquid-liquid critical point (LLCP) in deeply supercooled water leading to the co-existence of two phases: a high- and low-density liquid (HDL and LDL). While the thermodynamics associated with this putative LLCP has been well characterised through numerical simulations, the dynamical properties of these two phases close to the critical point remain much less understood. In this work, we investigate their dynamical and spectroscopic  features using machine-learning interatomic potentials (MLIPs). Dynamical analyses using the van-Hove correlation function, reveal that LDL exhibits very sluggish and heterogeneous molecular mobility, in contrast to the faster and more homogeneous dynamics of HDL. Infrared absorption (IR) spectra further show clear vibrational distinctions between LDL and HDL, in particular in the far IR region between $400-1000 \, \textrm{cm}^{-1}$. Together, these findings provide new dynamical fingerprints that clarify the microscopic behavior of supercooled water and offer valuable guidance for experimental efforts aimed at detecting the long-sought liquid–liquid transition.
\end{abstract}

\maketitle

\section{\label{sec:introduction} Introduction}
Water is one of the most essential liquids on Earth, playing a crucial role in numerous physical, chemical and biological processes \cite{Ball2007}. This central importance stems from its remarkably complex, unusual, and unique physical properties often referred to as anomalies, since they deviate significantly from the behavior typically observed in other liquids \cite{Handle2017}. While these anomalous properties manifest across a broad range of temperatures and pressures, they become particularly pronounced under supercooled conditions, when water remains in the liquid state below its freezing point by preventing ice nucleation \cite{Debenedetti2003}. One leading hypothesis proposed to explain these anomalies is that supercooled water can exist in two distinct liquid phases: a low-density liquid (LDL) and a high-density liquid (HDL) \cite{Poole1992,Poole2005}. In this scenario, water's anomalies are thought to originate from the presence of a liquid–liquid critical point (LLCP) in the supercooled region of its phase diagram, which is believed to drive its unusual thermodynamic and dynamic behavior \cite{Stanley1980}. 

The LLCP hypothesis was originally proposed based on molecular dynamics (MD) simulations using classical force fields \cite{Poole1992} and has since been supported by extensive theoretical studies \cite{Sciortino2011,Palmer2013,Debenedetti2020}. Recently, the advent of machine learning and particularly the development of machine-learning interatomic potentials (MLIPs) \cite{Behler2007,Bartk2010,Weinan2018} has opened new avenues for investigating this scenario with quantum mechanical accuracy at a nearly empirical-potential cost. Initial studies showed that MLIPs trained on high-quality density functional theory (DFT) data, gave further credence to the liquid–liquid transition in supercooled water \cite{Gartner2020,Gartner2022}. Notably, a recent breakthrough employed an MLIP trained on coupled-cluster-level reference data, providing the strongest evidence to date for the LLCP and precisely locating it within water’s phase diagram \cite{Sciortino2025}. Beyond this specific application, MLIPs have generally enabled the systematic exploration of water’s structural and thermodynamic properties more broadly, overcoming the limitations of direct ab initio molecular dynamics \cite{Zhang2021,Tisi2021,Piaggi2021,Malosso2022,Gomez2024}.

On the experimental front, although direct observation of the LLCP remains challenging due to rapid crystallization, indirect evidence continues to accumulate through various measurements \cite{Mallamace2007,Sellberg2014,Kim2017,Kim2020}. Yet, a clear-cut experimental confirmation of this elusive transition is still lacking. Recent studies have employed advanced spectroscopic techniques to probe water's properties deep within the so-called "no-man’s land". In the recent years, Kringle et al. \cite{Kringle2020,Kringle2023} investigated structural transformations in supercooled water films transiently heated across a temperature range spanning from deeply supercooled to ambient conditions, by analyzing the evolution of the infrared (IR) spectrum. Remarkably, they were able to decompose the IR signal into distinct contributions associated with low- and high-density liquid motifs. From the theoretical side, however, there is currently a lack of results supporting these experimental findings, as  IR spectra of supercooled water have not yet been systematically investigated. This gap largely stems from the fact that the field has long been dominated by classical water models, which are inadequate for accurately describing high-frequency vibrational spectroscopy \cite{Rossi2014,Domina2025}.

The complex thermodynamic phase diagram of water under supercooling is also manifested in various dynamical anomalies  \cite{Sciortino1996,Gallo1996,Jedlovszky2011,Saito2024}. These include, for example, deviations from the Stokes–Einstein relationship \cite{Kumar2006,Dehaoui2015,MonterodeHijes2018} and the emergence of dynamical heterogeneities \cite{Stirnemann2012,Kuo2021}. Such behaviors reflect the increasingly correlated spatial and temporal dynamics that arise as water is cooled into the deeply supercooled regime, particularly near the proposed liquid–liquid critical point. These anomalies have reinforced the view that structural fluctuations between low- and high-density local environments are intimately linked to changes in the dynamics \cite{Netz2002,Tanaka2025}, although a complete microscopic understanding remains elusive. Despite these insights, most prior work relies on empirical force fields, and the extent to which such anomalies are perturbed under more accurate quantum mechanical descriptions remains an open question. In this regard, we recently demonstrated, using MLIP simulations, that the LDL phase exhibits persistent, long-lived polarization fluctuations over hundreds of nanoseconds, which are reminescent of ferroelectric features\cite{Malosso2024}. Notably, the timescales associated with the underlying rotational dynamics were found to be highly sensitive to the level of theory, showing significant differences between empirical models versus the MLIP. These findings motivate a more detailed characterization of the microscopic dynamics across the supercooled phases, particularly between high-density and low-density liquids, as a key step toward understanding the subtle interplay between the thermodynamic structure and dynamics in water.

In this work, we build on these earlier results and probe deeper into the dynamical and spectroscopic properties of low- and high-density liquid water through MD simulations based on a MLIP trained on accurate DFT data \cite{Zhang2021}. We employ the Deep Potential framework \cite{Zeng2023} in combination with maximally localized Wannier functions (MLWFs) \cite{Marzari1997,Souza2001} to study water at the Strongly Constrained and Appropriately Normed (SCAN) \cite{Sun2015} meta-GGA level of theory. We first analyze dynamical properties such as the mean squared displacement, the self part of the van Hove correlation function, and indicators of dynamical heterogeneity in order to characterize the contrasting mobility and collective behavior of HDL and LDL. We then turn to predicting infrared spectra from our simulations and present the first theoretical spectra of LDL and HDL water, revealing clear spectroscopic fingerprints that distinguish the two liquids, particularly in the far-infrared region where collective intermolecular motions dominate. These computational predictions of the low-frequency IR spectra should provoke and help in interpreting future experiments in the field.

\section{\label{sec:methods}COMPUTATIONAL DETAILS}

We perform MD simulations of bulk water using the SCAN-based MLIP model developed in \cite{Zhang2021} leveraging \texttt{DeePMD-kit} \cite{Zeng2023} and using the \texttt{LAMMPS} package \cite{Thompson2022}. Two initial simulations have been carried out under distinct thermodynamic conditions to investigate the  LDL and HDL phase of supercooled water. Specifically, $500 \, \mathrm{ns}$-long NPT simulations of 1536 water molecules were run at $235 \, \textrm{K}$ and $3600 \, \textrm{bar}$ for the HDL phase, and at $235 \, \textrm{K}$ and $2800 \, \textrm{bar}$ for the LDL phase, conditions at which each phase is found to be stable with respect to the other \cite{Gartner2022}. Temperature and pressure were controlled using the Nosé–Hoover thermostat and barostat, with relaxation times of $0.5 \, \textrm{ps}$ and $5 \, \textrm{ps}$, respectively. From these simulations, the resulting equilibrium density of $1.02 \, \textrm{kg}\cdot \textrm{m}^{-3}$ for the LDL phase and $1.17 \, \textrm{kg}\cdot \textrm{m}^{-3}$ for the HDL phase were used to initialize subsequent NVT simulations. The temperature was controlled using the Bussi–Donadio–Parrinello thermostat \cite{Bussi2007} with a coupling constant of $8 \, \textrm{ps}$ at the target temperature of $235 \, \textrm{K}$ for both phases. For each phase, five independent NVT trajectories were generated: $100 \, \textrm{ns}$-long for the HDL phase and $400 \,\textrm{ns}$-long for the LDL phase. In addition to the supercooled liquid phases, a $50 \, \textrm{ns}$-long NVT simulation was run at $330 \, \textrm{K}$, targeting ambient liquid water.

The NVT simulations for both LDL and HDL were analyzed to characterize the dynamical properties of the systems. A primary quantity of interest is the Mean Squared Displacement (MSD), defined as:
\begin{equation}
    \textrm{MSD}(t) = \frac{1}{N}\sum_i\langle | \mathbf{r}_i\left(t\right)-\mathbf{r}_i\left(0\right) |^2 \rangle,
    \label{eq:msd}
\end{equation}
where $\mathbf{r}_i(t)$ is the atomic position of the i-th particle at time $t$ and $N$ is the number of particles. In the long-time limit, for systems that follow Fickian diffusion, the MSD grows linearly with time:
 \begin{equation}
     \textrm{MSD}(t) \sim 6Dt,
     \label{eq:diff}
 \end{equation}
where $D$ is the diffusion coefficient and can be obtained by analyzing the slope of the MSD curve as a function of time. To gain a deeper understanding of the microscopic dynamics of supercooled water, we also computed the self part of the  van Hove correlation function \cite{VanHove1954}, $G_S(r,t)$, defined as:
\begin{equation}
    G_S(r,t) = \frac{1}{N} \langle \sum_i \delta(\mathbf{r}_i(t)-\mathbf{r}_i(0)-\mathbf{r})\rangle\SB{.}
    \label{eq:vanhove}
\end{equation}
$G_S(r,t)$ quantifies the probability density  of a particle initially located at a given position $\mathbf{r}_i(0)$ and displaced by a vector $\mathbf{r}$ after a time interval $t$, thereby characterizing the single-particle displacement dynamics over time and providing detailed insight into the distribution of molecular displacements by going beyond average quantities like the mean squared displacement. 
In the context of supercooled liquids and glass-forming systems, the self van Hove function is widely used to identify and quantify dynamical heterogeneity, a hallmark of slow and nonuniform relaxation dynamics \cite{Donati1998,Hopkins2010}. In particular, deviations from Gaussian behavior are indicative of the coexistence of mobile and dynamically arrested particles in the liquid \cite{Bhowmik2018, Haddadian2017}. Thus, analysis of $G_S(r,t)$ serves as a sensitive probe of the spatial and temporal fluctuations in molecular motion, complementing more conventional dynamical observables \cite{Shinohara2020}.

To probe the high-frequency dynamics of the systems, $50 \, \textrm{ns}$ segments were extracted from each NVT simulation and used to compute the IR spectra, $\alpha(\omega)n(\omega)$ as the Fourier transform of the total dipole $\mathbf{M}$ time correlation function
\begin{equation}
    \alpha(\omega)n(\omega)=  \frac{2\pi \omega^2 \beta}{3cV} \int_{-\infty}^{+\infty} dt e^{-i\omega t} \langle \mathbf{M}(t)\mathbf{M}(0) \rangle.
    \label{eq:IR}
\end{equation}
where $\omega$ is the angular frequency,
c the speed of light, V the volume of the simulation box and $\beta = (k_B T)^{-1}$ the inverse temperature.
The total dipole $\mathbf{M}(t)=\sum\mathbf{\mu}_i(t)$ is computed as the sum of the instantaneous dipole moments $\mathbf{\mu}(t)$ of individual molecules. These molecular dipoles are predicted by a deep neural network trained on reference dipoles derived from SCAN-DFT calculations, using MLWF decomposition of the electronic density, as previously conducted from Ref. \cite{Malosso2024}. Further insight into the microscopic origins of the infrared response can be gained by decomposing the IR spectrum in Eq. \eqref{eq:IR} into its self and cross correlation terms. These represent, respectively, the intramolecular and intermolecular contributions to the dipole–dipole time correlation function \cite{Chen2008}. This yields:
\begin{multline}
\alpha(\omega)n(\omega) = \frac{2\pi \omega^2 \beta}{3cV} \int_{-\infty}^{+\infty} dt\, e^{-i\omega t} \times \\ \left[ 
\sum_i \langle \boldsymbol{\mu}_i(t) \cdot \boldsymbol{\mu}_i(0) \rangle  +
 \sum_{i \neq j} \langle \boldsymbol{\mu}_i(t) \cdot \boldsymbol{\mu}_j(0) \rangle 
\right]
\label{eq:dec-IR}
\end{multline}
where the first term represents the self (intra-molecular) contribution, and the second term accounts for the cross (inter-molecular) dipole correlations. Similar types of decomposition have previously been conducted based on DFT-based ab initio molecular dynamics simulations to study collective modes manifested in the low-frequency parts of the IR spectrum \cite{Heyden2010,Hlzl2021}.

\section{\label{sec:introduction} Results}

\begin{figure}[!t]
\centering
  \includegraphics[width=0.48\textwidth]
  {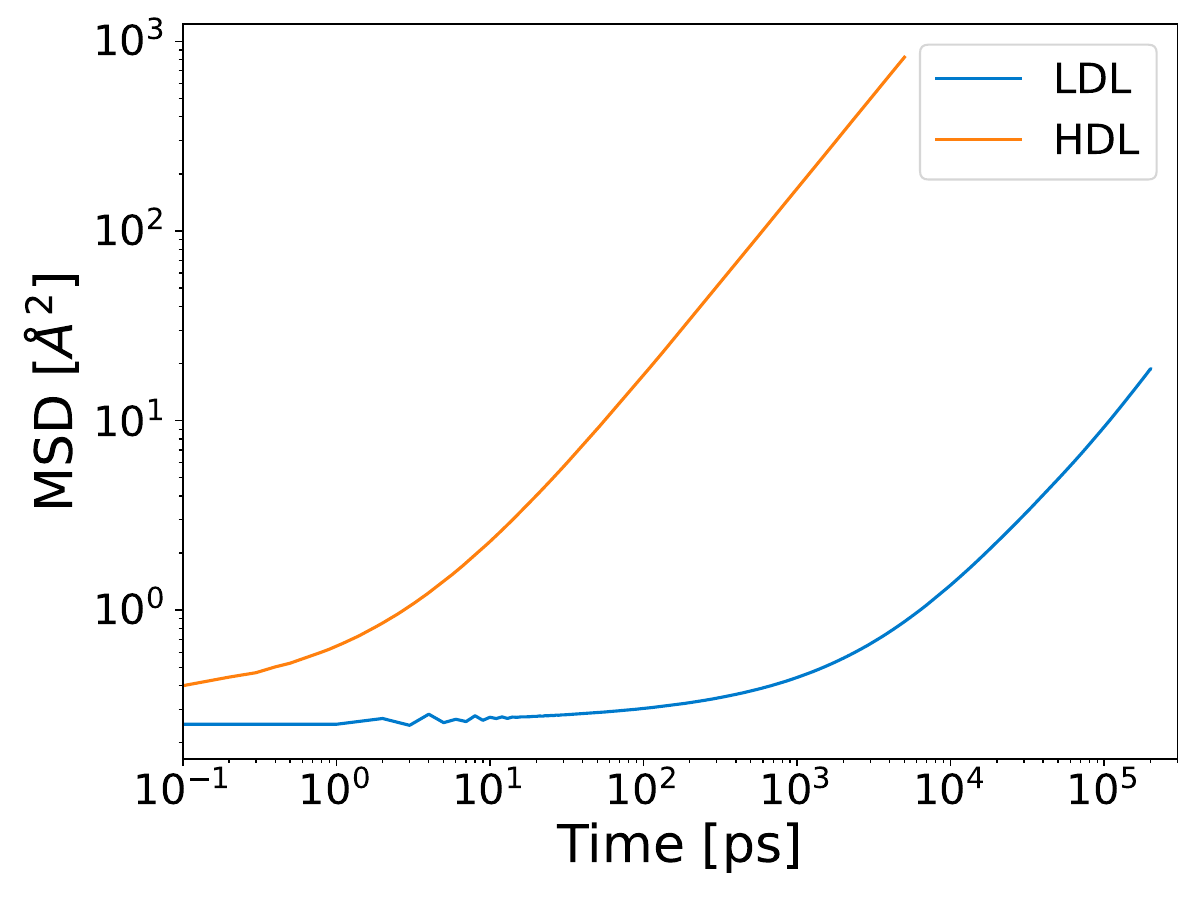}
\caption{Mean squared displacement as a function of time for LDL (blue) and HDL (orange) water. While both phases exhibit diffusive behavior at long times, the HDL phase shows significantly faster molecular mobility.}
\label{fig:msd}
\end{figure}

We begin our analysis of the dynamical properties of supercooled water by computing the mean squared displacement for both systems, as defined in Eq. \eqref{eq:msd}. The resulting curves are shown in Fig. \ref{fig:msd}, with LDL and HDL represented in blue and orange, respectively. Due to the sampling rate of $1 \, \textrm{ps}^{-1}$, the initial ballistic regime ($\mathrm{MSD}\propto t^2$) is not properly resolved in our data. Instead, both MSD curves begin with a plateau, reflecting transient caging where the diffusion of molecules is temporarily blocked by their neighbors, consistent with previous observations from classical empirical simulations of water\cite{Sciortino1996,Gallo1996}. At longer times, the curves transition to a linear regime, characteristic of normal diffusive behavior. During the caging period, the molecules undergo localized, constrained motions, and are thus unable to diffuse freely. In HDL, the trapping lasts up to few picoseconds, after which the molecules escape the cages and the system enters the diffusive regime. In contrast, the caging effect persists much longer in LDL, with molecules remaining trapped for up to several nanoseconds before the onset of diffusion, highlighting the stronger confinement of molecules in the LDL phase compared to HDL. 

The diffusion coefficient was then computed from Eq. \eqref{eq:diff} by fitting the linear regime of the MSD curve. The associated uncertainty was estimated as the standard deviation across the five independent NVT trajectories for each system. We obtained a diffusion coefficient of $D = 2.77 \pm 0.20 \times 10^{-3} \, \text{\AA}^2/\mathrm{ps}$ for HDL, and $D = 1.7 \pm 0.1 \times 10^{-5} \, \text{\AA}^2/\mathrm{ps}$ for LDL, confirming the markedly suppressed mobility in the low-density phase. For reference, the same machine-learning model yields a diffusion coefficient of $0.2 \, \text{\AA}^2/\mathrm{ps}$ for ambient liquid water, in close agreement with the benchmark value from direct ab initio molecular dynamics simulations \cite{Chen2017} and in good agreement with experimental values of $0.187 \text{\AA}^2/\mathrm{ps}$ \cite{Mills1973}. For LDL, it is worth noting that the system does not fully reach the diffusive regime within the simulation timescale, and thus the extracted diffusion coefficient may underestimate the true long-time limit.

\begin{figure*}[!ht]
  \centering
  \includegraphics[width=0.3\textwidth]{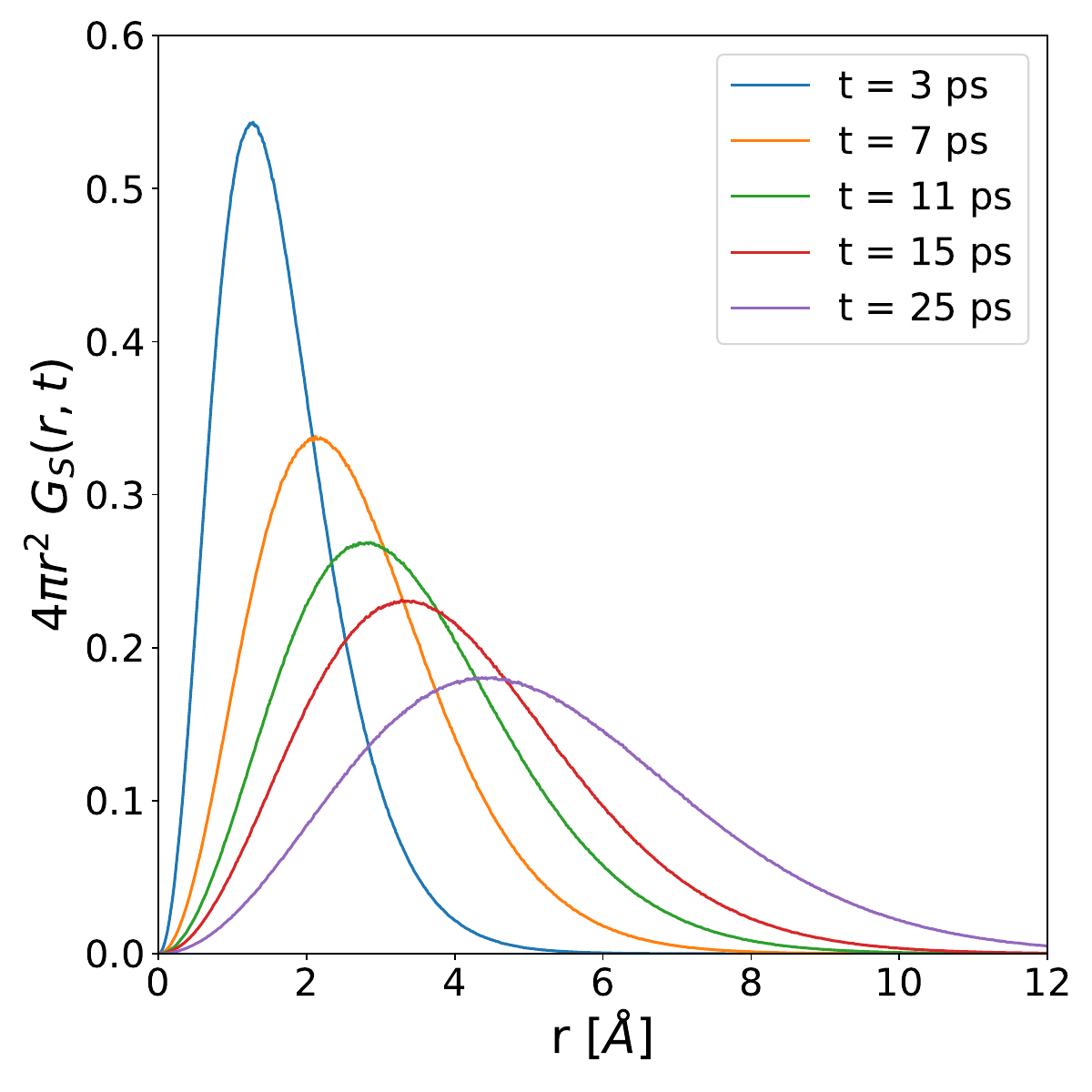}
  \includegraphics[width=0.3\textwidth]{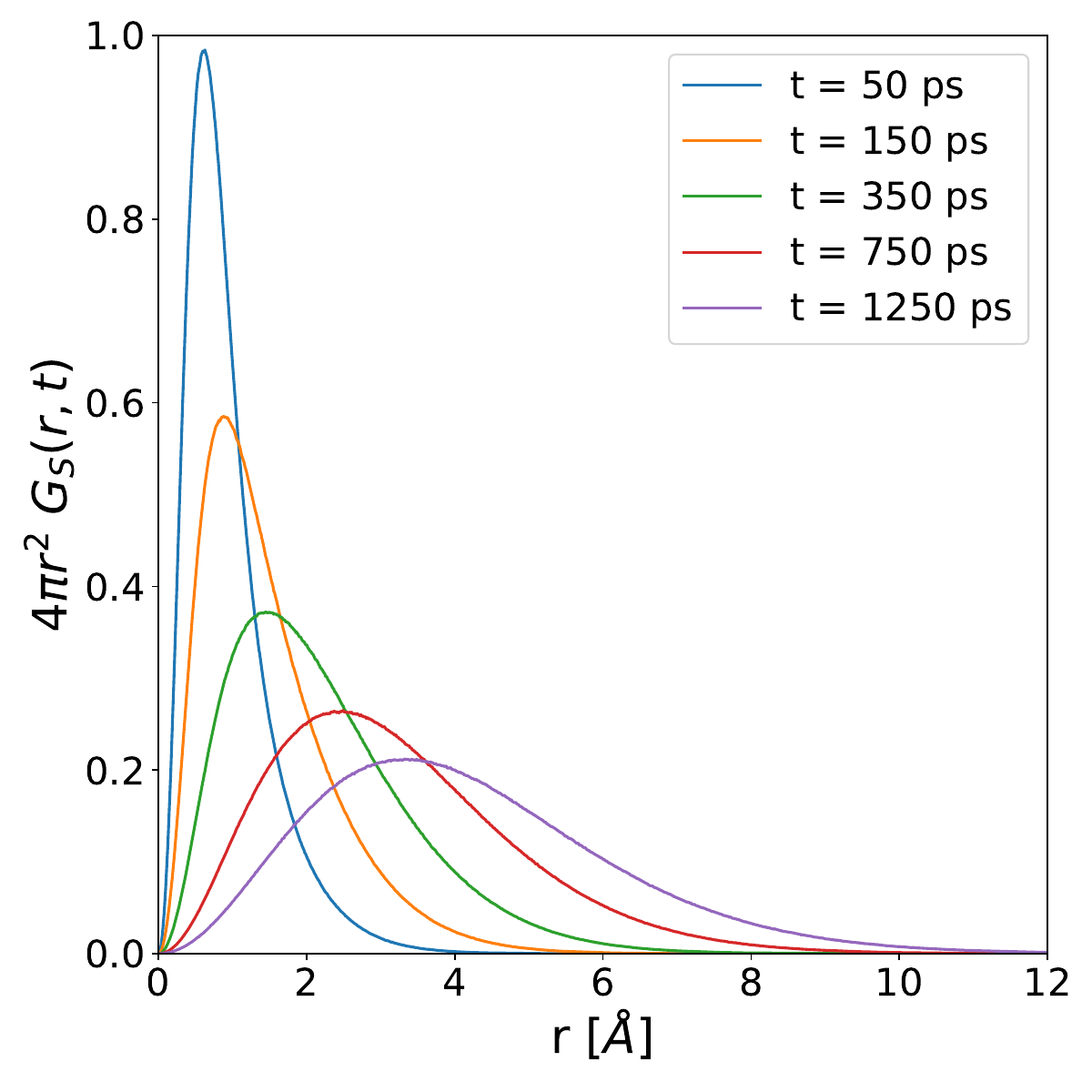}
  \includegraphics[width=0.3\textwidth]{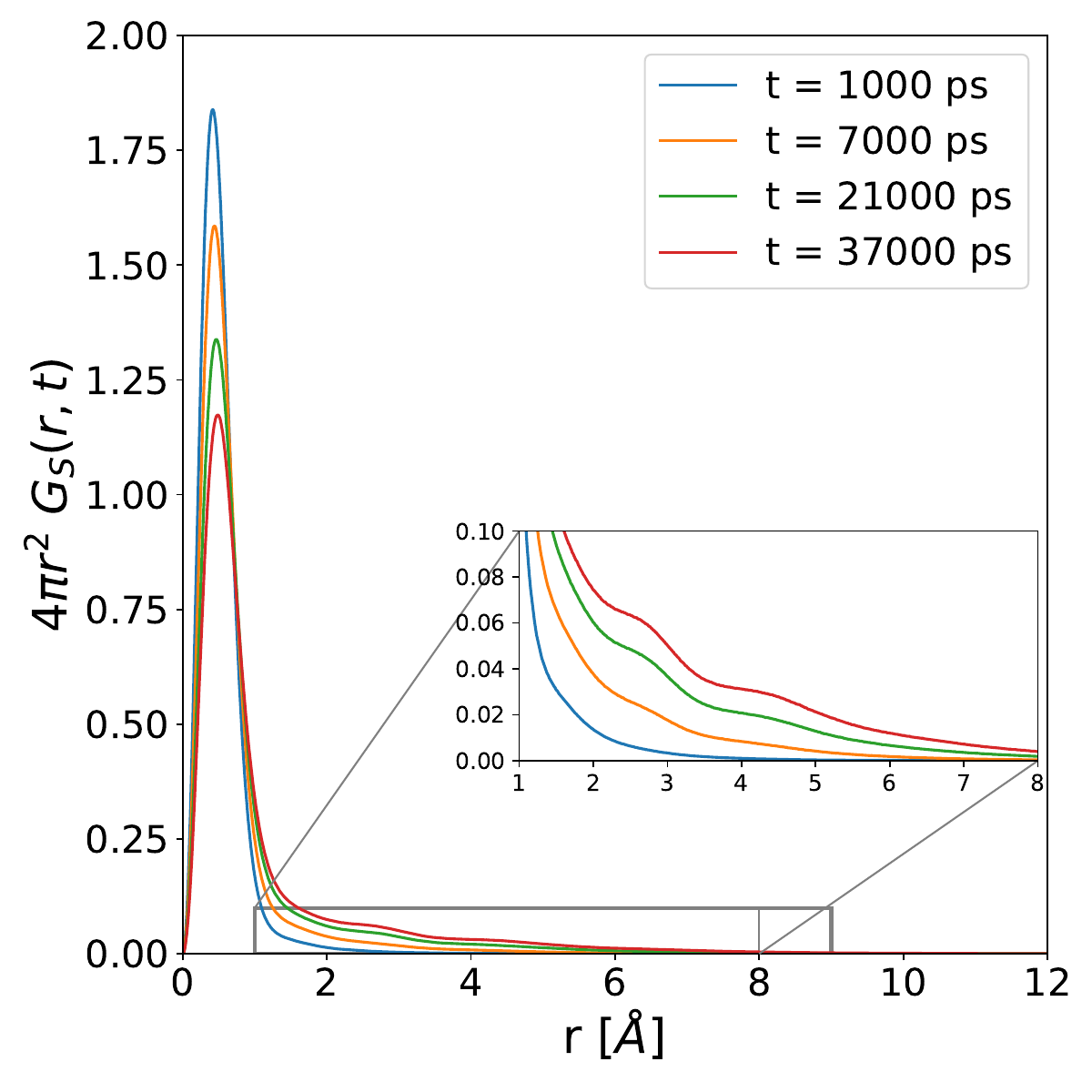}
  \caption{Self part of the van Hove correlation function $ G_S(r,t)$ for ambient liquid water (\emph{left}), HDL (\emph{middle}) and HDL (\emph{right}) at selected time intervals. Liquid water and HDL phase exhibits a smooth shift in the peak over time. In contrast, LDL shows a slower and more complex evolution, with pronounced tail and different peaks at intermediate timescales, indicative of dynamic heterogeneity.}
  \label{fig:vanhove-comparison}
\end{figure*}

To gain deeper insight into the microscopic dynamics underlying these diffusivities and to capture features that could be lost in averaging, we computed the scalar self part of the Van Hove correlation function $G_s(r,t)$, as given from Eq. \eqref{eq:vanhove}. 
The results of this analysis are shown in Fig. \ref{fig:vanhove-comparison} as $4\pi r^2 G_S(r,t)$ for liquid water at ambient conditions, HDL, and LDL. Different colors indicate the time intervals at which the Van Hove function was evaluated. Comparing ambient water and HDL, we observe that, aside from the systematically slower dynamics in HDL, their Van Hove functions  are qualitatively very similar with each displaying a single peak that shifts to larger displacements over time, reflecting relatively homogeneous dynamics.
On the other hand, LDL deviates significantly from the other two, developing fat tails with subtle shoulders for times larger than $\sim 5 \, \textrm{ns}$.  In addition, while at room temperature and in HDL,  the peak position shifts as expected from a diffusive process, in LDL instead, the tails develop over time, but with the peak positions only changing by $0.05 \,\textrm{\AA}$ after $37 \, \textrm{ns}$. These results suggest that while some particles remain relatively localized, populating the first peak in the Van Hove function, others move significantly farther over the same time interval, giving rise to exponential tails in the Van Hove function. Such features are indicative of spatially heterogeneous dynamics and are not present in the HDL phase. To further support this interpretation, we computed the non-Gaussianity parameter $\alpha_2(t)$~\cite{Rahman1964}, which quantifies deviations of the Van Hove correlation function, $G_S(r,t)$, from a purely Gaussian distribution. Our analysis shows that, for ambient water and HDL, $\alpha_2(t)$ decays toward zero—indicating Gaussian behavior, within tens and hundreds of picoseconds, respectively. In stark contrast, the LDL phase exhibits persistent non-Gaussian behavior, with a value of $\alpha_2(t = 40\,\text{ns}) \approx 4$. A more detailed analysis is presented in Fig. 1 of the Supporting Material.

\begin{figure}[t]
\centering
\includegraphics[width=0.48\textwidth]{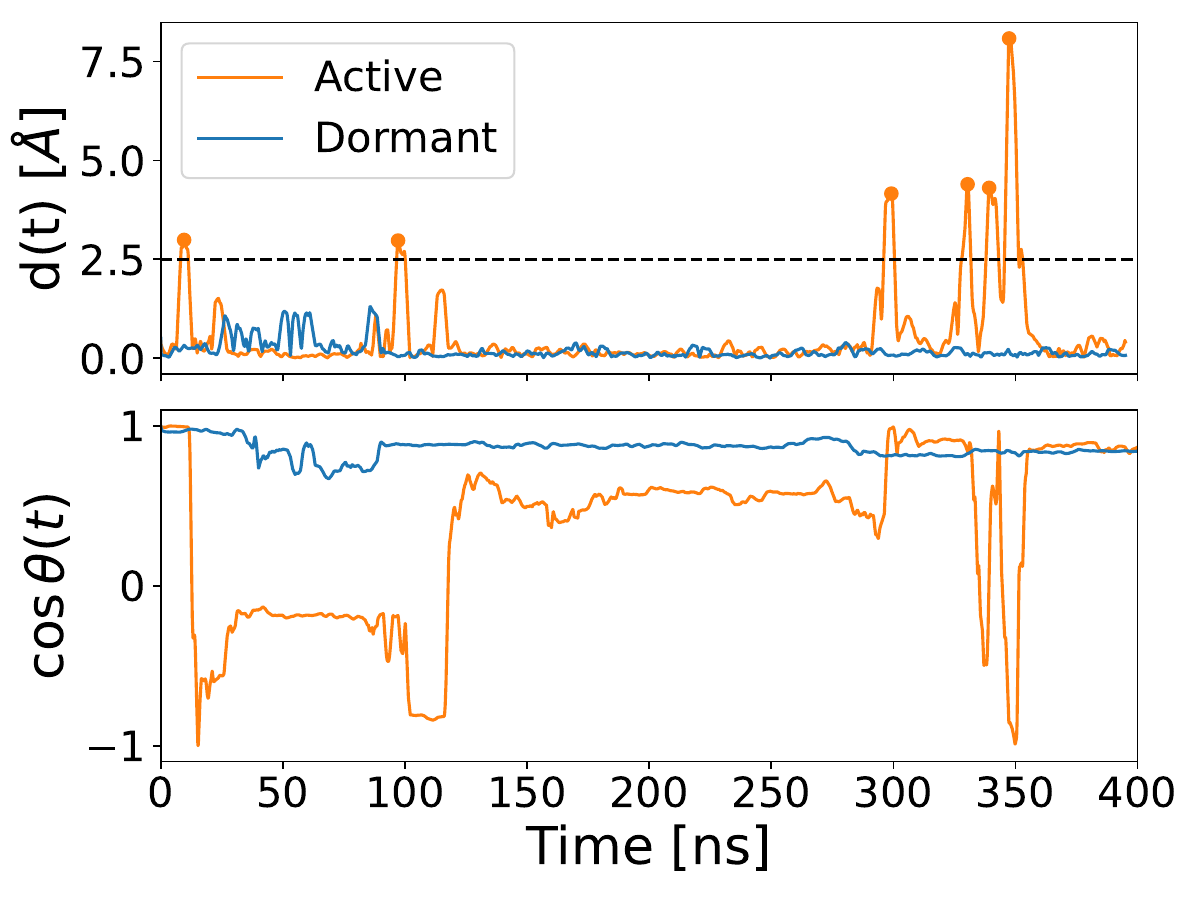}
\includegraphics[width=0.48\textwidth]{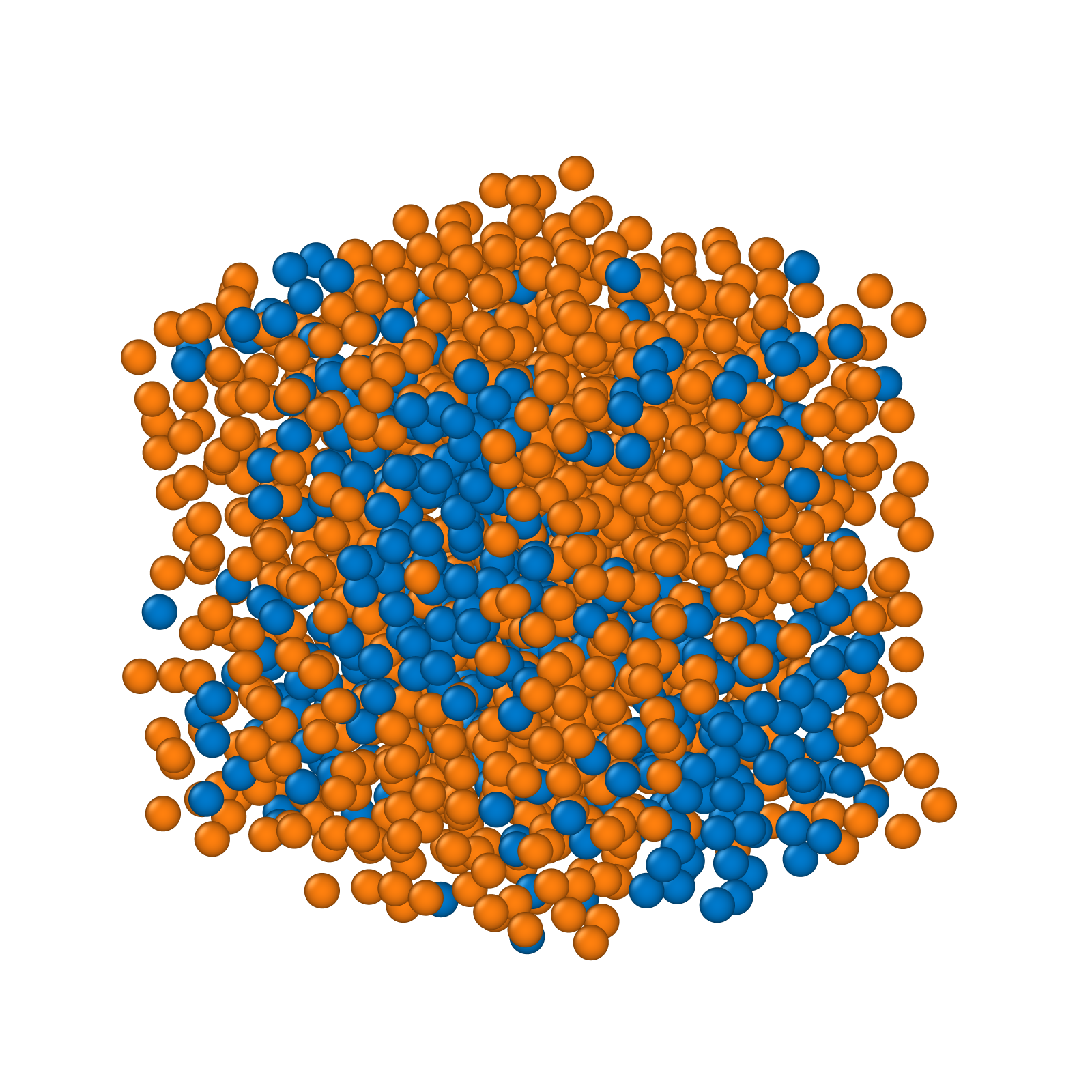}
\caption{(a) Displacement time series $d(t)$ for a dormant (blue) and an active (orange) water molecule. The dashed line marks the jump threshold at $2.5 \, \text{\AA}$. Peaks above this threshold correspond to jump-like events. (b) Snapshot of the simulation box. Molecules classified as \emph{dormant} are shown in blue; \emph{active} molecules are shown in orange. Roughly 30 \% of the molecules are dynamically arrested, indicating spatial heterogeneity in mobility. }
\label{fig:dormant_active}
\end{figure}

Upon visual inspection of our LDL trajectories, there appeared to be events where some water molecules remained trapped for very long times (100s of nanoseconds), while others exhibited sudden, large bursts of motion. To corroborate these descriptive qualitative observations, inspired by a technique developed in Ref. \cite{OffeiDanso2023} to detect abrupt discontinuities in the reorientational dynamics of water molecules, we applied a similar protocol to analyze particle dynamics and identify jump-like translational motions and dynamical inhomogeneities in LDL water. The time series of the atomic positions of each molecule (tracked via its oxygen atom) was first processed to reduce noise and extract meaningful dynamical features, by applying a second-order low-pass digital butterworth filter with a cutoff frequency set at $1$ GHz. Additionally, we refined the trajectories by applying a mean filter using the smooth function with a span of 2000 data points (equivalent to a window of $2 \, \textrm{ns}$). 

Displacements over a lag time of $\Delta t=5 \, \mathrm{ns}$ were then calculated as the modulus of the difference between smoothed positions separated by this interval, $d_i(t) = \left\| \tilde{\mathbf{r}_i}(t + \Delta t) - \tilde{\mathbf{r}_i(t)} \right\|$, where $\tilde{\mathbf{r}}$ is the filtered time-series of the atomic position of the i-th molecule. The resulting displacement data were subsequently analyzed to identify peaks associated with abrupt translational motions. Jumps were defined as displacements exceeding $2.5 \, \text{\AA}$, a threshold slightly shorter than the typical hydrogen bond length, and were counted for each water molecule accordingly. With this criterion, we classified molecules based on their dynamical behavior: \emph{dormant} molecules are those that do not undergo any jump (i.e., zero displacements above $2.5 \, \text{\AA}$ over the entire $400 \, \textrm{ns}$-long simulation), while active molecules are those that exhibit at least one such jump. This definition allows us to distinguish dynamically arrested species from those contributing to translational dynamics. 

In Fig. \ref{fig:dormant_active} a), we report the displacement time series for two representative molecules: one classified as dormant and the other as active. The active molecule exhibits several sharp peaks crossing the jump threshold, each marked with a circle in the figure, while the dormant one shows only sub-threshold fluctuations. In addition, Figure \ref{fig:dormant_active} b) presents the time series of the dipole angle change with respect to its initial value. As one can observe, the translational jumps are strongly coupled to orientational rearrangements: the dormant molecule maintains a nearly constant orientation throughout the simulation, while the active one undergoes significant rotational motion simultaneously with its translational jumps.

Our analyses reveal that approximately 30\% of water molecules on average remain dynamically arrested over the full simulation timescale. These statistics were obtained by an analysis of all the 5 independently run simulations of LDL amounting to $2 \, \mu s$. In contrast no such events are observed in HDL - no water molecules are identified as dormant. An instantaneous snapshot of the LDL liquid for one particular frame is shown in Fig. \ref{fig:dormant_active} c), where dormant molecules are highlighted in blue and active molecules in orange. The spatial distribution indicates partial clustering of dormant species, suggesting that the dynamical heterogeneity in the LDL phase involves a collective hydrodynamic effect. Our findings echo a recent work by Saito \cite{Saito2024}, who investigated the slowdown of dynamics in supercooled water using classical empirical models. By analyzing dynamic disorder and survival probabilities, the author identified that cooperative rearrangements play a crucial role in facilitating translational jumps. Notably, they observed that at temperatures where the LDL-like state becomes dominant, jump events become increasingly intermittent and spatially correlated due to enhanced cooperativity in molecular motion. This picture is strongly supported by our results, which reveal persistent non-Gaussian dynamics and jump-like translational and rotational behavior in LDL, consistent with a dynamically arrested, heterogeneous environment governed by long-lived collective reorientations.

\begin{figure}[t]
\centering
\includegraphics[width=0.4\textwidth]{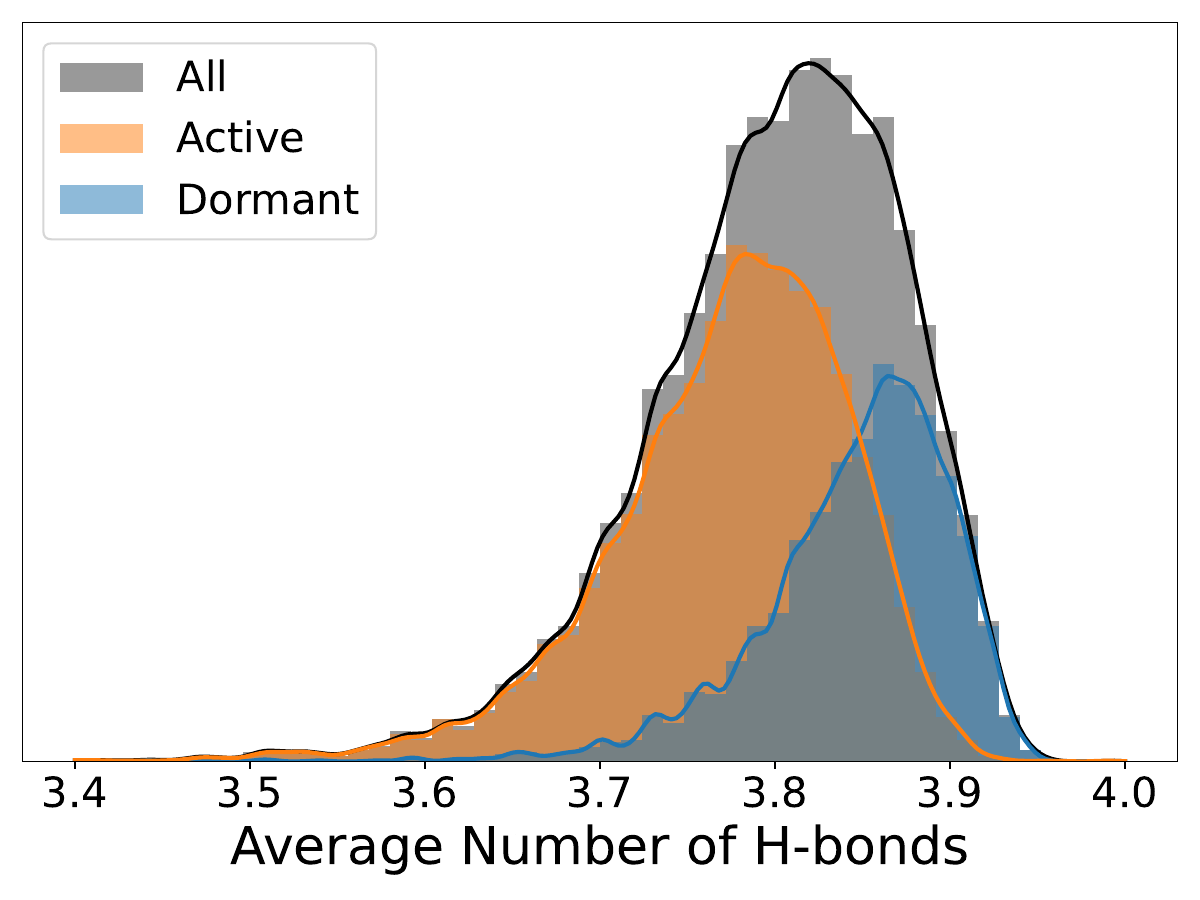}
\caption{Histogram of the time-averaged number of hydrogen bonds per molecule. The total distribution is decomposed into contributions from dynamically \emph{dormant} (blue) and \emph{active} (orange) molecules. Active molecules tend to be undercoordinated and exhibit a broader distribution with a tail toward fewer hydrogen bonds.}
\label{fig:hbond_hist}
\end{figure}

A natural question that emerges from the preceding analyses is whether the dynamical heterogeneity observed is manifested in any structural parameters of the water hydrogen-bond network. We thus performed an analysis of structural defects based on the average number of hydrogen bonds formed by each molecule. It is indeed known that defective molecules act as catalytic sites, promoting mobility and diffusion \cite{Sciortino1991}. For each water molecule, we computed the time-averaged number of hydrogen bonds over the $400 \, \textrm{ns}$ trajectory. A geometrical criterion akin to that proposed by Luzar and Chandler\cite{Luzar1996} was used. This criterion considers two water molecules to be hydrogen bonded when the distance between the donating and accepting oxygen is within $3.0 \, \text{\AA}$ and the angle formed by the bond vector between the donating hydrogen and oxygen and the bond vector between the donating and accepting oxygen is less than $30^\circ$. Note that the distance criterion is slightly adjusted from what is typically used at room temperature owing to the different structuring as noted previously by us\cite{Malosso2024}. The resulting distribution was then decomposed into contributions from dynamically dormant and active molecules as previously discussed. As shown in Fig. \ref{fig:hbond_hist}, the distributions of hydrogen bonds for both dormant (blue) and active (orange) molecules largely overlap, indicating that the number of hydrogen bonds alone is not sufficient to fully distinguish between dynamically arrested and mobile species. Nevertheless, a clear statistical trend emerges: the distribution for dormant molecules is shifted toward higher hydrogen bond counts, while the distribution for active molecules displays a broader shape and a long tail extending to lower hydrogen bond counts. This observation confirms that molecules with fewer hydrogen bonds are more likely to undergo jump-like displacements, but that mobility cannot be attributed to local structural defects alone. Instead, it highlights that dynamical activity likely arises from a subtle interplay of local environment, hydrogen bonding fluctuations, and collective rearrangements, rather than from isolated undercoordinated defects. However, it is worth noting that in our current scheme, a molecule is classified as \emph{active} if it exhibits a jump motion at least once during the $400 \, \textrm{ns}$ trajectory. Thus, one might observe bigger distinctions in the number of hydrogen bonds if one factors also the number of dynamical events involved in the observation time window.
%(XXX IF A WATER WAS ACIVE ONCE IN THE 400NS IT GETS LUMPED AS ACIVE RIGHT? XXX PERHAPS WE SHOULD CLARIFY THIS POINT IN THE STATISTICS WE CONSTRUCT - IT DOESN'T DISTINGUISH WHICH MAY EXPLAIN THE OVERLAP OF THE H-BONDING?)

\begin{figure}[t]
\centering
  \includegraphics[width=0.48\textwidth]
  {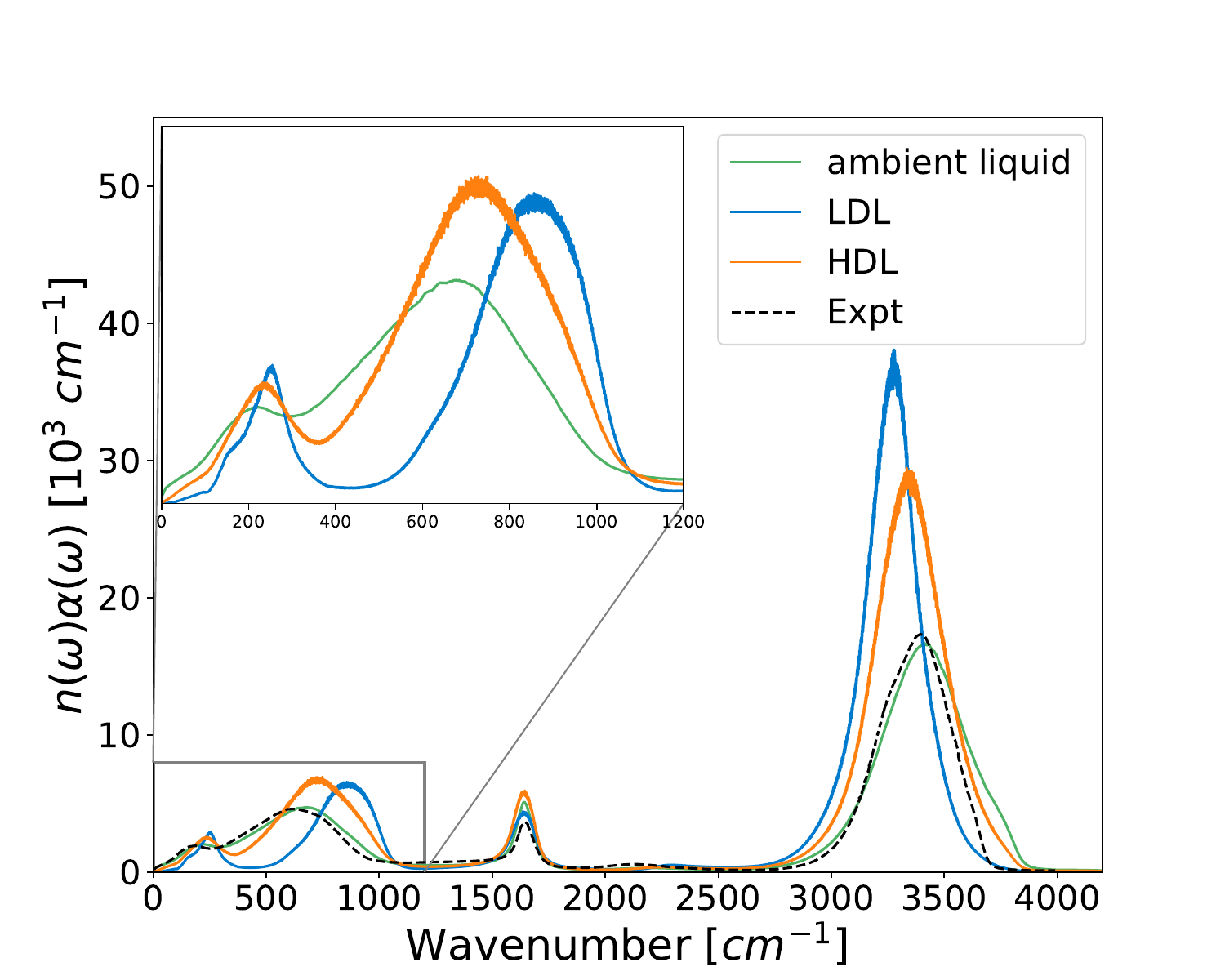}

  \caption{Infrared absorption spectra of ambient, LDL, and HDL water compared with experimental data. Simulated IR spectra computed using dipole moments at the SCAN-DFT level for ambient liquid water, LDL, and HDL. The spectra reveal distinct features across the frequency range, with particularly marked differences in far-IR region. The comparison with experimental data \cite{Bertie1996} validates the accuracy of the SCAN functional and the MLIP in capturing spectroscopic features of liquid water.}
\label{fig:ir_spectra}
\end{figure}

\begin{figure}[t]
\centering
  \includegraphics[width=0.48\textwidth]
  {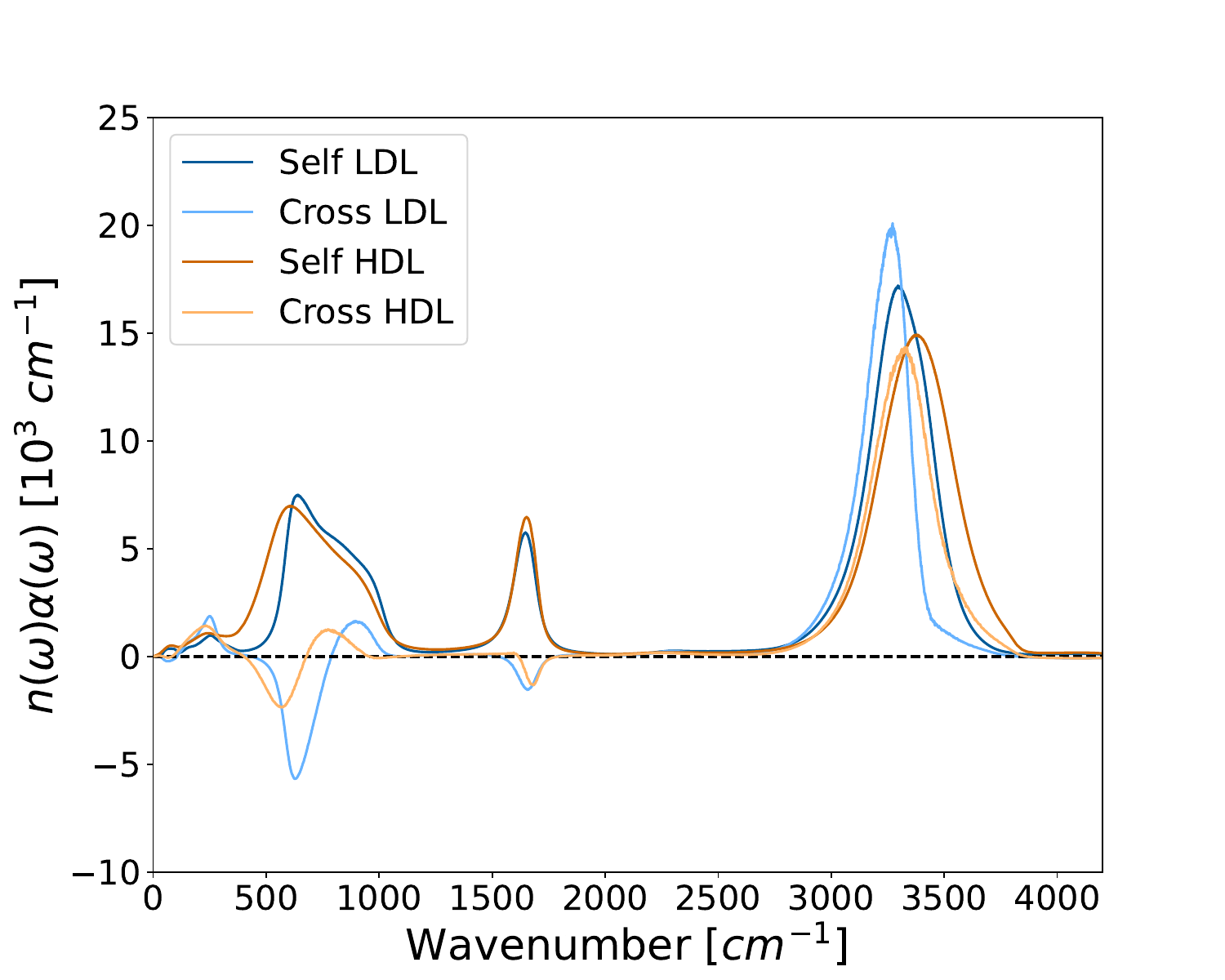}

\caption{Decomposition of the IR absorption spectra of HDL and LDL into self (intra-molecular) and cross (inter-molecular) contributions. The low-frequency region (particularly in the libration band ($400\text{--}1000 \, \text{cm}^{-1}$) is primarily governed by the cross term, highlighting the role of collective dipolar correlations.}
\label{fig:ir_decomposition}
\end{figure}

%XXX FOR THE FIGURE ABOVE SHOWING THE CROSS CAN YOU PLEASE MAKE THEM IN DASHED LINES XXX (\CM{NO, too noisy and too many points, you won't see anything, and it's already filthered...})

If the LDL phase features dynamical heterogeneity, it is interesting to explore if and how this would be reflected in the vibrational IR spectra. In a recent work, we proposed through an examination of the polarization fluctuations, that the LDL phase exhibits ferroelectric-like features\cite{Malosso2024}. Izzo and co-workers have further hypothesized the existence of a ferroelectric phase transition occurring concurrently with the liquid–liquid phase transition \cite{Izzo2024}. We computed the IR of HDL, LDL and ambient temperature liquid water. The results are reported in Fig. \ref{fig:ir_spectra}. Experimental values for the IR spectrum of ambient liquid water are also reported as taken from Ref. \cite{Bertie1996}, confirming the ability of the SCAN-DFT MLIP to correctly reproduce the vibrational properties of liquid water. 

As expected and consistent with previous studies, the high-frequency region of the spectrum, corresponding to the OH stretching vibrations ($\sim 3400\,\text{cm}^{-1}$), red-shifted for the LDL phase by $40 \, \textrm{cm}^{-1}$ compared to HDL indicating that the hydrogen-bond strengths of the water molecules in LDL are stronger. On the other hand, there is substantial overlap in the spectra for the two liquids indicating that the O-H oscillators immersed in local environments that create similar electric fields. Moreover, substantial differences appear when comparing the OH stretch band of the supercooled systems to that of ambient liquid water, with both LDL and HDL exhibiting red-shifted peaks. This shift reflects the stronger and more structured hydrogen bonding present in the supercooled phases relative to the ambient liquid.

In contrast, the low-frequency region reveals pronounced differences, particularly in the libration band ($400\text{--}1000 \, \text{cm}^{-1}$), which is sensitive to hindered rotational motion. The LDL spectrum shows a sharper, blue-shifted libration peak (by approximately $140 \, \textrm{cm}^{-1}$) compared to HDL, which is broader in terms of its FWHM. This difference reflects the more rigid and structured hydrogen-bond network in LDL, which imposes stronger constraints on rotational motion. Conversely, the broader librational response in HDL indicates enhanced rotational freedom, consistent with its more disordered environment. Additionally, a distinct spectral feature appears around $200 \, \text{cm}^{-1}$, commonly attributed to intermolecular hydrogen-bond stretching modes. This peak is more intense and slightly blue-shifted in LDL, reflecting stiffer hydrogen-bond cages and stronger intermolecular dipole–dipole correlations. This further supports the interpretation that LDL is characterized by more collective dynamics.

To better understand the origin of the spectral differences between LDL and HDL, we decomposed the IR spectra into self and cross contributions as in Eq. \eqref{eq:dec-IR} described earlier in the methods section. Comparing the self and the cross correlation contributions to the librational band reveals two important aspects. Firstly, the cross correlation spectra consists of both positive and negative components corresponding to correlated and anti-correlated dipole fluctuations at those frequencies respectively. Secondly, the blue shift observed in LDL relative to HDL in the low-frequency region, including both the libration and hydrogen-bond stretching peaks, arises predominantly from the cross component. The cross correlations probe the coupling of the motion of a water molecule with its surrounding environment and is thus particularly sensitive to the structure and strength of the hydrogen-bond network. In LDL, where hydrogen bonds are stronger and the local environment is more rigid, these intermolecular correlations are enhanced, resulting in a blue-shift in both the libration and the hydrogen-bond stretch modes. All in all, the more pronounced differences in the low-frequency region, contrasted with the large overlap of of the high-frequency OH stretching band, suggest that the key distinctions between the two supercooled phases arise from collective, large-scale spatial and temporal correlations, rather than from variations in local molecular structure. Recently, Roke and co-workers have introduced a new technique called correlated vibrational spectroscopy  (CVS) which can be used to extract interacting versus non-interacting molecular contributions to the vibrational response, by separating self and cross-correlation spectra \cite{Flr2024}. Extending such experimental strategies to supercooled water could offer a promising route to directly observe the low-frequency features discussed here and provide new probes for supercooled water.

\section{Conclusions}

In this work, we investigated the dynamical and spectroscopic properties of HDL and LDL water using long-timescale MD simulations based on MLIP at the SCAN-DFT level. Our results reveal striking differences in both properties probed for the two phases. From a dynamical perspective, LDL exhibits significantly slower and more heterogeneous molecular motion compared to HDL. Analysis of the self part of the van Hove correlation function shows that while HDL displays homogeneous, diffusive dynamics, LDL is characterized by pronounced dynamical arrest and heterogeneous mobility, with a substantial fraction of molecules remaining trapped over hundreds of nanoseconds. A classification of molecules into \emph{active} and \emph{dormant} subsets further confirmed the presence of spatially heterogeneous dynamics in LDL, with active molecules undergoing intermittent, jump-like displacements. Further structural analysis indicates that such dynamical heterogeneity cannot be fully attributed to static defects in the hydrogen-bond network, suggesting that mobility arises from a more complex interplay of hydrogen-bond fluctuations and collective rearrangements consistent with previous studies\cite{Saito2018,Kuo2021}.

Complementing the dynamical analysis, we computed the infrared absorption spectra of LDL, HDL, and ambient liquid water. We observe a pronounced blue shift and narrowing of the libration band in LDL, indicating enhanced rigidity and stronger intermolecular correlations in its hydrogen-bond network. A decomposition of the IR spectra into self and cross dipole contributions demonstrates that this spectral shift arises primarily from enhanced cross-correlations in LDL, underscoring the collective origin of its vibrational response.

Taken all together, our results provide microscopic dynamical and spectroscopic fingerprints that distinguish LDL and HDL. These findings not only advance our theoretical understanding of water’s anomalous behavior in the deeply supercooled regime but also offer guidance for interpreting spectroscopic signatures in future experiments aiming to detect the elusive liquid–liquid transition. One other key take-home message within this context is that future experiments should carefully look at the librational band of supercooled water as there appears to be some key fingerprints reflecting the differences in dynamics between the two liquids.

\section*{Supplementary Material}
The Supplementary Material for this publication includes additional analyses supporting the main text.

\begin{acknowledgments}
This work was partially supported by the European Commission through the MaX Centre of Excellence for supercomputing applications (grant number 101093374) and by the Italian MUR, through the PRIN project ARES (grant number 2022W2BPCK) and the Italian National Centre from HPC, Big Data, and Quantum Computing (grant number CN00000013), founded within the Next Generation EU initiative. A.H acknowledges the funding
received by the European Research Council (ERC) under the European Union’s Horizon 2020 research and innovation programme (grant number 101043272 - HyBOP).
\end{acknowledgments}

\section*{Data Availability Statement}
Owing to their large sizes, all MD trajectories are stored on a dedicated local server and are available from the corresponding author upon request.

\nocite{*}
\bibliography{aipsamp}% Produces the bibliography via BibTeX.

\end{document}